\documentstyle[12pt,a4,epsfig]{article}
\newcommand{\neutr}{\chi}

\def \stop {\tilde{\mathrm{t}}}
\def \sbot {\tilde{\mathrm{b}}}
\def \snu {\tilde{\nu}}
\def \neu {\chi}
\def \deltm {\Delta m}
\def \thetamix  {\theta_{\mathrm{\tilde{t}}}}
\def \thetab  {\theta_{\mathrm{\tilde{b}}}}

\def \ggqq {\gamma\gamma \rightarrow \rm{q\bar{q}}}
\def \ggtt {\mathrm \gamma\gamma \rightarrow \tau^{+}\tau^{-}}

\def \qqg  {\rm{q\bar{q}}(\gamma)}

\def \ww   {\mathrm WW}
\def \zz   {\mathrm Z\gamma^{*}}

\def \ewnu {\mathrm We\nu}
\def \rts  {\sqrt{s}}
\def \pt   {p_{\mathrm{t}}}

\def \mvis {M_{\rm{vis}}}
\def \thscat {\theta_{\mathrm{scat}}}

\def \stopr {\mathrm \tilde{t}_{R}}
\def \stopl {\mathrm \tilde{t}_{L}}

\def \gev  { \, \mathrm{GeV}/\it{c}^{\mathrm{2}}}

\begin{document}
\pagestyle{empty}

\bigskip
\begin{flushright} 
\vspace{1cm}
\end{flushright} 
\vspace{2cm}
\begin{center}
  \mathversion{bold}
  {\LARGE\bf Scalar quark searches  in ${\mathrm e^+ e^-}$ collisions
 at $\sqrt{{\mathrm s}}=181-184$ GeV }
  \mathversion{normal}
  \vskip 1.2cm
  {\bf The ALEPH Collaboration}\\ 
  \vskip 1.2cm
  {\bf Abstract}
\end{center}
\medskip
%
%
%
\small

Searches for scalar top, scalar bottom and degenerate scalar quarks have been 
performed with data collected with the ALEPH detector at LEP.  
The data sample consists of 57~$\mathrm{pb}^{-1}$ taken at $\rts$ =
181--184~GeV. 
No evidence for scalar top, scalar bottom or degenerate
scalar quarks was found in the channels $\stop \rightarrow \mathrm{c}\neu$, 
$\stop \rightarrow \mathrm{b}\ell\snu$, $\sbot \rightarrow
\mathrm{b}\neu$, and $\mathrm{\tilde{q}} \rightarrow \mathrm{q}\neu$.
From the channel $\stop \rightarrow \mathrm{c}\neu$ a limit of 
74~$\gev$ has been set on the scalar top quark mass, independent of the 
mixing angle. This limit assumes a mass difference between the
$\stop$ and the $\neu$ in the range 10--40~$\gev$. From the channel
$\stop \rightarrow \mathrm{b}\ell\snu$ the mixing-angle-independent scalar top
limit is 82~$\gev$,
 assuming  $m_{\mathrm{\tilde{t}}}-m_{\tilde{\nu}}$ $>$ 10~$\gev$. 
From the channel $\sbot \rightarrow \mathrm{b}\neu$, a limit of 79~$\gev$ 
has been set on the mass of the supersymmetric partner of the
 left-handed state of the bottom quark. This limit is valid 
 for $m_{\mathrm{\tilde{b}}}-m_{\neu}$ $>$ 10~$\gev$. From the channel 
$\mathrm{\tilde{q}} \rightarrow \mathrm{q}\neu$, a
limit of 87~$\gev$ has been set on the mass of supersymmetric partners
of light quarks assuming five degenerate flavours and the production of
both ``left-handed'' and ``right-handed'' squarks.  
This limit is valid for $m_{\mathrm{\tilde{q}}}-m_{\neu}$ $>$ 5~$\gev$. 
\vspace{1cm}
\begin{center}
\vspace{1cm}
\end{center}

\vfill
\pagebreak

 \small
\pagestyle{empty}
\newpage
\small
%
%
\newlength{\saveparskip}
\newlength{\savetextheight}
\newlength{\savetopmargin}
\newlength{\savetextwidth}
\newlength{\saveoddsidemargin}
\newlength{\savetopsep}
\setlength{\saveparskip}{\parskip}
\setlength{\savetextheight}{\textheight}
\setlength{\savetopmargin}{\topmargin}
\setlength{\savetextwidth}{\textwidth}
\setlength{\saveoddsidemargin}{\oddsidemargin}
\setlength{\savetopsep}{\topsep}
%
%
\setlength{\parskip}{0.0cm}
\setlength{\textheight}{25.0cm}
\setlength{\topmargin}{-1.5cm}
\setlength{\textwidth}{16 cm}
\setlength{\oddsidemargin}{-0.0cm}
\setlength{\topsep}{1mm}
\pretolerance=10000
\centerline{\large\bf The ALEPH Collaboration}
\footnotesize
\vspace{0.5cm}
{\raggedbottom
\begin{sloppypar}
\samepage\noindent
R.~Barate,
D.~Buskulic,
D.~Decamp,
P.~Ghez,
C.~Goy,
S.~Jezequel,
J.-P.~Lees,
A.~Lucotte,
F.~Martin,
E.~Merle,
\mbox{M.-N.~Minard},
\mbox{J.-Y.~Nief},
B.~Pietrzyk
\nopagebreak
\begin{center}
\parbox{15.5cm}{\sl\samepage
Laboratoire de Physique des Particules (LAPP), IN$^{2}$P$^{3}$-CNRS,
F-74019 Annecy-le-Vieux Cedex, France}
\end{center}\end{sloppypar}
\vspace{2mm}
\begin{sloppypar}
\noindent
R.~Alemany,
G.~Boix,
M.P.~Casado,
M.~Chmeissani,
J.M.~Crespo,
M.~Delfino, 
E.~Fernandez,
M.~Fernandez-Bosman,
Ll.~Garrido,$^{15}$
E.~Graug\`{e}s,
A.~Juste,
M.~Martinez,
G.~Merino,
R.~Miquel,
Ll.M.~Mir,
P.~Morawitz,
I.C.~Park,
A.~Pascual,
I.~Riu,
F.~Sanchez
\nopagebreak
\begin{center}
\parbox{15.5cm}{\sl\samepage
Institut de F\'{i}sica d'Altes Energies, Universitat Aut\`{o}noma
de Barcelona, 08193 Bellaterra (Barcelona), E-Spain$^{7}$}
\end{center}\end{sloppypar}
\vspace{2mm}
\begin{sloppypar}
\noindent
A.~Colaleo,
D.~Creanza,
M.~de~Palma,
G.~Gelao,
G.~Iaselli,
G.~Maggi,
M.~Maggi,
S.~Nuzzo,
A.~Ranieri,
G.~Raso,
F.~Ruggieri,
G.~Selvaggi,
L.~Silvestris,
P.~Tempesta,
A.~Tricomi,$^{3}$
G.~Zito
\nopagebreak
\begin{center}
\parbox{15.5cm}{\sl\samepage
Dipartimento di Fisica, INFN Sezione di Bari, I-70126 Bari, Italy}
\end{center}\end{sloppypar}
\vspace{2mm}
\begin{sloppypar}
\noindent
X.~Huang,
J.~Lin,
Q. Ouyang,
T.~Wang,
Y.~Xie,
R.~Xu,
S.~Xue,
J.~Zhang,
L.~Zhang,
W.~Zhao
\nopagebreak
\begin{center}
\parbox{15.5cm}{\sl\samepage
Institute of High-Energy Physics, Academia Sinica, Beijing, The People's
Republic of China$^{8}$}
\end{center}\end{sloppypar}
\vspace{2mm}
\begin{sloppypar}
\noindent
D.~Abbaneo,
U.~Becker,
\mbox{P.~Bright-Thomas},
D.~Casper,
M.~Cattaneo,
V.~Ciulli,
G.~Dissertori,
H.~Drevermann,
R.W.~Forty,
M.~Frank,
F.~Gianotti,
R.~Hagelberg,
J.B.~Hansen,
J.~Harvey,
P.~Janot,
B.~Jost,
I.~Lehraus,
P.~Maley,
P.~Mato,
A.~Minten,
L.~Moneta,$^{20}$
N.~Qi,
A.~Pacheco,
F.~Ranjard,
L.~Rolandi,
D.~Rousseau,
D.~Schlatter,
M.~Schmitt,$^{1}$
O.~Schneider,
W.~Tejessy,
F.~Teubert,
I.R.~Tomalin,
M.~Vreeswijk,
H.~Wachsmuth
\nopagebreak
\begin{center}
\parbox{15.5cm}{\sl\samepage
European Laboratory for Particle Physics (CERN), CH-1211 Geneva 23,
Switzerland}
\end{center}\end{sloppypar}
\vspace{2mm}
\begin{sloppypar}
\noindent
Z.~Ajaltouni,
F.~Badaud
G.~Chazelle,
O.~Deschamps,
A.~Falvard,
C.~Ferdi,
P.~Gay,
C.~Guicheney,
P.~Henrard,
J.~Jousset,
B.~Michel,
S.~Monteil,
\mbox{J-C.~Montret},
D.~Pallin,
P.~Perret,
F.~Podlyski,
J.~Proriol,
P.~Rosnet
\nopagebreak
\begin{center}
\parbox{15.5cm}{\sl\samepage
Laboratoire de Physique Corpusculaire, Universit\'e Blaise Pascal,
IN$^{2}$P$^{3}$-CNRS, Clermont-Ferrand, F-63177 Aubi\`{e}re, France}
\end{center}\end{sloppypar}
\vspace{2mm}
\begin{sloppypar}
\noindent
J.D.~Hansen,
J.R.~Hansen,
P.H.~Hansen,
B.S.~Nilsson,
B.~Rensch,
A.~W\"a\"an\"anen
\begin{center}
\parbox{15.5cm}{\sl\samepage
Niels Bohr Institute, 2100 Copenhagen, DK-Denmark$^{9}$}
\end{center}\end{sloppypar}
\vspace{2mm}
\begin{sloppypar}
\noindent
G.~Daskalakis,
A.~Kyriakis,
C.~Markou,
E.~Simopoulou,
A.~Vayaki
\nopagebreak
\begin{center}
\parbox{15.5cm}{\sl\samepage
Nuclear Research Center Demokritos (NRCD), GR-15310 Attiki, Greece}
\end{center}\end{sloppypar}
\vspace{2mm}
\begin{sloppypar}
\noindent
A.~Blondel,
\mbox{J.-C.~Brient},
F.~Machefert,
A.~Roug\'{e},
M.~Rumpf,
R.~Tanaka,
A.~Valassi,$^{6}$
H.~Videau
\nopagebreak
\begin{center}
\parbox{15.5cm}{\sl\samepage
Laboratoire de Physique Nucl\'eaire et des Hautes Energies, Ecole
Polytechnique, IN$^{2}$P$^{3}$-CNRS, \mbox{F-91128} Palaiseau Cedex, France}
\end{center}\end{sloppypar}
\vspace{2mm}
\begin{sloppypar}
\noindent
E.~Focardi,
G.~Parrini,
K.~Zachariadou
\nopagebreak
\begin{center}
\parbox{15.5cm}{\sl\samepage
Dipartimento di Fisica, Universit\`a di Firenze, INFN Sezione di Firenze,
I-50125 Firenze, Italy}
\end{center}\end{sloppypar}
\vspace{2mm}
\begin{sloppypar}
\noindent
R.~Cavanaugh,
M.~Corden,
C.~Georgiopoulos,
T.~Huehn,
D.E.~Jaffe
\nopagebreak
\begin{center}
\parbox{15.5cm}{\sl\samepage
Supercomputer Computations Research Institute,
Florida State University,
Tallahassee, FL 32306-4052, USA $^{13,14}$}
\end{center}\end{sloppypar}
\vspace{2mm}
\begin{sloppypar}
\noindent
A.~Antonelli,
G.~Bencivenni,
G.~Bologna,$^{4}$
F.~Bossi,
P.~Campana,
G.~Capon,
F.~Cerutti,
V.~Chiarella,
G.~Felici,
P.~Laurelli,
G.~Mannocchi,$^{5}$
F.~Murtas,
G.P.~Murtas,
L.~Passalacqua,
M.~Pepe-Altarelli
\nopagebreak
\begin{center}
\parbox{15.5cm}{\sl\samepage
Laboratori Nazionali dell'INFN (LNF-INFN), I-00044 Frascati, Italy}
\end{center}\end{sloppypar}
\vspace{2mm}
\begin{sloppypar}
\noindent
M.~Chalmers,
L.~Curtis,
A.W.~Halley,
J.G.~Lynch,
P.~Negus,
V.~O'Shea,
C.~Raine,
J.M.~Scarr,
K.~Smith,
P.~Teixeira-Dias,
A.S.~Thompson,
E.~Thomson,
J.J.~Ward
\nopagebreak
\begin{center}
\parbox{15.5cm}{\sl\samepage
Department of Physics and Astronomy, University of Glasgow, Glasgow G12
8QQ,United Kingdom$^{10}$}
\end{center}\end{sloppypar}
\pagebreak
\begin{sloppypar}
\noindent
O.~Buchm\"uller,
S.~Dhamotharan,
C.~Geweniger,
G.~Graefe,
P.~Hanke,
G.~Hansper,
V.~Hepp,
E.E.~Kluge,
A.~Putzer,
J.~Sommer,
K.~Tittel,
S.~Werner,
M.~Wunsch
\nopagebreak
\begin{center}
\parbox{15.5cm}{\sl\samepage
Institut f\"ur Hochenergiephysik, Universit\"at Heidelberg, D-69120
Heidelberg, Germany$^{16}$}
\end{center}\end{sloppypar}
\vspace{2mm}
\begin{sloppypar}
\noindent
R.~Beuselinck,
D.M.~Binnie,
W.~Cameron,
P.J.~Dornan,$^{12}$
M.~Girone,
S.~Goodsir,
E.B.~Martin,
N.~Marinelli,
A.~Moutoussi,
J.~Nash,
J.K.~Sedgbeer,
P.~Spagnolo,
M.D.~Williams
\nopagebreak
\begin{center}
\parbox{15.5cm}{\sl\samepage
Department of Physics, Imperial College, London SW7 2BZ,
United Kingdom$^{10}$}
\end{center}\end{sloppypar}
\vspace{2mm}
\begin{sloppypar}
\noindent
V.M.~Ghete,
P.~Girtler,
E.~Kneringer,
D.~Kuhn,
G.~Rudolph
\nopagebreak
\begin{center}
\parbox{15.5cm}{\sl\samepage
Institut f\"ur Experimentalphysik, Universit\"at Innsbruck, A-6020
Innsbruck, Austria$^{18}$}
\end{center}\end{sloppypar}
\vspace{2mm}
\begin{sloppypar}
\noindent
C.K.~Bowdery,
P.G.~Buck,
P.~Colrain,
G.~Crawford,
A.J.~Finch,
F.~Foster,
G.~Hughes,
R.W.L.~Jones,
A.N.~Robertson,
M.I.~Williams
\nopagebreak
\begin{center}
\parbox{15.5cm}{\sl\samepage
Department of Physics, University of Lancaster, Lancaster LA1 4YB,
United Kingdom$^{10}$}
\end{center}\end{sloppypar}
\vspace{2mm}
\begin{sloppypar}
\noindent
I.~Giehl,
C.~Hoffmann,
K.~Jakobs,
K.~Kleinknecht,
M.~Kr\"ocker,
H.-A.~N\"urnberger,
G.~Quast,
B.~Renk,
E.~Rohne,
H.-G.~Sander,
P.~van~Gemmeren,
C.~Zeitnitz,
T.~Ziegler
\nopagebreak
\begin{center}
\parbox{15.5cm}{\sl\samepage
Institut f\"ur Physik, Universit\"at Mainz, D-55099 Mainz, Germany$^{16}$}
\end{center}\end{sloppypar}
\vspace{2mm}
\begin{sloppypar}
\noindent
J.J.~Aubert,
C.~Benchouk,
A.~Bonissent,
G.~Bujosa,
J.~Carr,$^{12}$
P.~Coyle,
A.~Ealet,
D.~Fouchez,
O.~Leroy,
F.~Motsch,
P.~Payre,
M.~Talby,
A.~Sadouki,
M.~Thulasidas,
A.~Tilquin,
K.~Trabelsi
\nopagebreak
\begin{center}
\parbox{15.5cm}{\sl\samepage
Centre de Physique des Particules, Facult\'e des Sciences de Luminy,
IN$^{2}$P$^{3}$-CNRS, F-13288 Marseille, France}
\end{center}\end{sloppypar}
\vspace{2mm}
\begin{sloppypar}
\noindent
M.~Aleppo, 
M.~Antonelli,
F.~Ragusa
\nopagebreak
\begin{center}
\parbox{15.5cm}{\sl\samepage
Dipartimento di Fisica, Universit\`a di Milano e INFN Sezione di
Milano, I-20133 Milano, Italy.}
\end{center}\end{sloppypar}
\vspace{2mm}
\begin{sloppypar}
\noindent
R.~Berlich,
V.~B\"uscher,
G.~Cowan,
H.~Dietl,
G.~Ganis,
G.~L\"utjens,
C.~Mannert,
W.~M\"anner,
\mbox{H.-G.~Moser},
S.~Schael,
R.~Settles,
H.~Seywerd,
H.~Stenzel,
W.~Wiedenmann,
G.~Wolf
\nopagebreak
\begin{center}
\parbox{15.5cm}{\sl\samepage
Max-Planck-Institut f\"ur Physik, Werner-Heisenberg-Institut,
D-80805 M\"unchen, Germany\footnotemark[16]}
\end{center}\end{sloppypar}
\vspace{2mm}
\begin{sloppypar}
\noindent
J.~Boucrot,
O.~Callot,
S.~Chen,
M.~Davier,
L.~Duflot,
\mbox{J.-F.~Grivaz},
Ph.~Heusse,
A.~H\"ocker,
A.~Jacholkowska,
M.~Kado,
D.W.~Kim,$^{2}$
F.~Le~Diberder,
J.~Lefran\c{c}ois,
L.~Serin,
E.~Tournefier,
\mbox{J.-J.~Veillet},
I.~Videau,
D.~Zerwas
\nopagebreak
\begin{center}
\parbox{15.5cm}{\sl\samepage
Laboratoire de l'Acc\'el\'erateur Lin\'eaire, Universit\'e de Paris-Sud,
IN$^{2}$P$^{3}$-CNRS, F-91898 Orsay Cedex, France}
\end{center}\end{sloppypar}
\vspace{2mm}
\begin{sloppypar}
\noindent
\samepage
P.~Azzurri,
G.~Bagliesi,$^{12}$
S.~Bettarini,
T.~Boccali,
C.~Bozzi,
G.~Calderini,
R.~Dell'Orso,
R.~Fantechi,
I.~Ferrante,
A.~Giassi,
A.~Gregorio,
F.~Ligabue,
A.~Lusiani,
P.S.~Marrocchesi,
A.~Messineo,
F.~Palla,
G.~Rizzo,
G.~Sanguinetti,
A.~Sciab\`a,
G.~Sguazzoni,
R.~Tenchini,
C.~Vannini,
A.~Venturi,
P.G.~Verdini
\samepage
\begin{center}
\parbox{15.5cm}{\sl\samepage
Dipartimento di Fisica dell'Universit\`a, INFN Sezione di Pisa,
e Scuola Normale Superiore, I-56010 Pisa, Italy}
\end{center}\end{sloppypar}
\vspace{2mm}
\begin{sloppypar}
\noindent
G.A.~Blair,
L.M.~Bryant,
J.T.~Chambers,
J.~Coles,
M.G.~Green,
T.~Medcalf,
P.~Perrodo,
J.A.~Strong,
J.H.~von~Wimmersperg-Toeller
\nopagebreak
\begin{center}
\parbox{15.5cm}{\sl\samepage
Department of Physics, Royal Holloway \& Bedford New College,
University of London, Surrey TW20 OEX, United Kingdom$^{10}$}
\end{center}\end{sloppypar}
\vspace{2mm}
\begin{sloppypar}
\noindent
D.R.~Botterill,
R.W.~Clifft,
T.R.~Edgecock,
S.~Haywood,
P.R.~Norton,
J.C.~Thompson,
A.E.~Wright
\nopagebreak
\begin{center}
\parbox{15.5cm}{\sl\samepage
Particle Physics Dept., Rutherford Appleton Laboratory,
Chilton, Didcot, Oxon OX11 OQX, United Kingdom$^{10}$}
\end{center}\end{sloppypar}
\vspace{2mm}
\begin{sloppypar}
\noindent
\mbox{B.~Bloch-Devaux},
P.~Colas,
B.~Fabbro,
G.~Fa\"\i f,
E.~Lan\c{c}on,$^{12}$
\mbox{M.-C.~Lemaire},
E.~Locci,
P.~Perez,
H.~Przysiezniak,
J.~Rander,
\mbox{J.-F.~Renardy},
A.~Rosowsky,
A.~Roussarie,
A.~Trabelsi,
B.~Vallage
\nopagebreak
\begin{center}
\parbox{15.5cm}{\sl\samepage
CEA, DAPNIA/Service de Physique des Particules,
CE-Saclay, F-91191 Gif-sur-Yvette Cedex, France$^{17}$}
\end{center}\end{sloppypar}
\vspace{2mm}
\begin{sloppypar}
\noindent
S.N.~Black,
J.H.~Dann,
H.Y.~Kim,
N.~Konstantinidis,
A.M.~Litke,
M.A. McNeil,
G.~Taylor
\nopagebreak
\begin{center}
\parbox{15.5cm}{\sl\samepage
Institute for Particle Physics, University of California at
Santa Cruz, Santa Cruz, CA 95064, USA$^{19}$}
\end{center}\end{sloppypar}
\pagebreak
\vspace{2mm}
\begin{sloppypar}
\noindent
C.N.~Booth,
S.~Cartwright,
F.~Combley,
M.S.~Kelly,
M.~Lehto,
L.F.~Thompson
\nopagebreak
\begin{center}
\parbox{15.5cm}{\sl\samepage
Department of Physics, University of Sheffield, Sheffield S3 7RH,
United Kingdom$^{10}$}
\end{center}\end{sloppypar}
\vspace{2mm}
\begin{sloppypar}
\noindent
K.~Affholderbach,
A.~B\"ohrer,
S.~Brandt,
J.~Foss,
C.~Grupen,
L.~Smolik,
F.~Stephan 
\nopagebreak
\begin{center}
\parbox{15.5cm}{\sl\samepage
Fachbereich Physik, Universit\"at Siegen, D-57068 Siegen, Germany$^{16}$}
\end{center}\end{sloppypar}
\vspace{2mm}
\begin{sloppypar}
\noindent
G.~Giannini,
B.~Gobbo,
G.~Musolino
\nopagebreak
\begin{center}
\parbox{15.5cm}{\sl\samepage
Dipartimento di Fisica, Universit\`a di Trieste e INFN Sezione di Trieste,
I-34127 Trieste, Italy}
\end{center}\end{sloppypar}
\vspace{2mm}
\begin{sloppypar}
\noindent
J.~Putz,
J.~Rothberg,
S.~Wasserbaech,
R.W.~Williams
\nopagebreak
\begin{center}
\parbox{15.5cm}{\sl\samepage
Experimental Elementary Particle Physics, University of Washington, WA 98195
Seattle, U.S.A.}
\end{center}\end{sloppypar}
\vspace{2mm}
\begin{sloppypar}
\noindent
S.R.~Armstrong,
A.P.~Betteridge,
E.~Charles,
P.~Elmer,
D.P.S.~Ferguson,
Y.~Gao,
S.~Gonz\'{a}lez,
T.C.~Greening,
O.J.~Hayes,
H.~Hu,
S.~Jin,
P.A.~McNamara III,
J.M.~Nachtman,$^{21}$
J.~Nielsen,
W.~Orejudos,
Y.B.~Pan,
Y.~Saadi,
I.J.~Scott,
J.~Walsh,
Sau~Lan~Wu,
X.~Wu,
G.~Zobernig
\nopagebreak
\begin{center}
\parbox{15.5cm}{\sl\samepage
Department of Physics, University of Wisconsin, Madison, WI 53706,
USA$^{11}$}
\end{center}\end{sloppypar}
}
\footnotetext[1]{Now at Harvard University, Cambridge, MA 02138, U.S.A.}
\footnotetext[2]{Permanent address: Kangnung National University, Kangnung,
Korea.}
\footnotetext[3]{Also at Dipartimento di Fisica, INFN Sezione di Catania,
Catania, Italy.}
\footnotetext[4]{Also Istituto di Fisica Generale, Universit\`{a} di
Torino, Torino, Italy.}
\footnotetext[5]{Also Istituto di Cosmo-Geofisica del C.N.R., Torino,
Italy.}
\footnotetext[6]{Supported by the Commission of the European Communities,
contract ERBCHBICT941234.}
\footnotetext[7]{Supported by CICYT, Spain.}
\footnotetext[8]{Supported by the National Science Foundation of China.}
\footnotetext[9]{Supported by the Danish Natural Science Research Council.}
\footnotetext[10]{Supported by the UK Particle Physics and Astronomy Research
Council.}
\footnotetext[11]{Supported by the US Department of Energy, grant
DE-FG0295-ER40896.}
\footnotetext[12]{Also at CERN, 1211 Geneva 23,Switzerland.}
\footnotetext[13]{Supported by the US Department of Energy, contract
DE-FG05-92ER40742.}
\footnotetext[14]{Supported by the US Department of Energy, contract
DE-FC05-85ER250000.}
\footnotetext[15]{Permanent address: Universitat de Barcelona, 08208 Barcelona,
Spain.}
\footnotetext[16]{Supported by the Bundesministerium f\"ur Bildung,
Wissenschaft, Forschung und Technologie, Germany.}
\footnotetext[17]{Supported by the Direction des Sciences de la
Mati\`ere, C.E.A.}
\footnotetext[18]{Supported by Fonds zur F\"orderung der wissenschaftlichen
Forschung, Austria.}
\footnotetext[19]{Supported by the US Department of Energy,
grant DE-FG03-92ER40689.}
\footnotetext[20]{Now at University of Geneva, 1211 Geneva 4, Switzerland.}
\footnotetext[21]{Now at University of California at Los Angeles (UCLA),
Los Angeles, CA 90024, U.S.A.}
%
%
\setlength{\parskip}{\saveparskip}
\setlength{\textheight}{\savetextheight}
\setlength{\topmargin}{\savetopmargin}
\setlength{\textwidth}{\savetextwidth}
\setlength{\oddsidemargin}{\saveoddsidemargin}
\setlength{\topsep}{\savetopsep}
\normalsize
\newpage
\pagestyle{plain}
\setcounter{page}{1}

\pagestyle{plain}
\setcounter{page}{1}
\normalsize
\section{Introduction}

In the Minimal Supersymmetric extension of the Standard Model
 (MSSM)~\cite{SUSY}, each chirality state of the 
  Standard Model fermions has a scalar supersymmetric partner.
 The scalar quarks (squarks) 
$\tilde{\rm{q}}_{\rm{R}}$ and $\tilde{\rm{q}}_{\rm{L}}$ are the
 supersymmetric partners of the left-handed and right-handed
 quarks, respectively.
 They are weak interaction eigenstates
which can mix to form the mass eigenstates.
Since the size of the mixing is 
proportional to the mass of the Standard Model partner, the lighter scalar 
top (stop) could be
the lightest supersymmetric charged particle.
The stop mass eigenstates are obtained by a unitary transformation of the 
$\stopr$ and $\stopl$ fields, parametrised by the mixing 
angle $\thetamix$. 
The lighter stop 
is given by $\stop = \mathrm{\tilde{t}_L \cos{\thetamix}}
+ \mathrm{\tilde{t}_R \sin{\thetamix}}$, while  
the heavier stop is the orthogonal combination. The stop could be produced 
at LEP in pairs, $\rm{e^+ e^-} \to \stop \bar{\stop}$, via {\it s}-channel
 exchange of a virtual photon or a Z.

The searches for stops described here assume that all supersymmetric
particles except the lightest neutralino $\neu$ and possibly the sneutrino
$\snu$ are heavier than the stop. The conservation of R-parity is also 
assumed; this implies that the lightest supersymmetric particle (LSP) 
is stable. Under these assumptions,
the two dominant decay channels are $\stop \to \rm{c} \neu$ 
and $\stop \to\rm{b} \ell \tilde{\nu}$~\cite{Hikasa}.
The first decay can only proceed via loops and thus 
has a very small width, of the order of 0.01--1~eV~\cite{Hikasa}. 
The $\stop \to \rm{b} \ell \tilde{\nu}$ 
channel proceeds via a virtual chargino exchange 
and has a width of the order of 0.1--10~keV~\cite{Hikasa}. 
The latter decay dominates when it is kinematically allowed. 
The phenomenology of the scalar bottom (sbottom), the supersymmetric 
partner of the bottom quark, is similar to the phenomenology of the
stop. 
Assuming that the $\sbot$ is lighter than all supersymmetric particles except
the $\neu$, the $\sbot$ will decay as $\mathrm{\tilde{b} \rightarrow
b \neu}$.  Compared to the $\stop$ decays, the $\sbot$ decay has
a large width of the order of 10--100~MeV.
Direct searches for stops and sbottoms 
are performed in the stop decay channels $\stop \to \rm{c}\neu$ and
$\stop \to\rm{b} \ell \tilde{\nu}$
and in the sbottom decay channel $\sbot \to \rm{b} \neu$.
The results of these searches supersede the ALEPH results reported 
earlier for data collected at energies up to 
$\rts$ = 172~GeV~\cite{ALEPH_stop}. 
The D0 experiment~\cite{D0} has reported a lower limit on the stop mass 
of 85~${\mathrm GeV}/c^2$ for the decay into $\rm{c} \neutr$ and for
a mass difference between the $\stop$ and the 
$\chi$ larger than about 40~$\gev$.
Searches for $\stop \rightarrow \rm{c} \neutr$, 
$\stop\rightarrow\mathrm{b} \ell \tilde{\nu}$ and
$\sbot\rightarrow\rm{b}\neutr$ using data
collected at LEP at energies up to $\sqrt{s}$ = 172~$\mathrm{GeV}$ have also 
been performed by OPAL~\cite{OPAL}.

The supersymmetric partners of
the light quarks are generally expected in the MSSM to
be heavy, i.e., beyond the reach of LEP2, but their
   masses receive large negative corrections
   from gluino loops~\cite{doni}. 
The dominant decay mode is assumed to be $\tilde{\rm{q}} \to \rm{q}\neu$.
Limits are set on the production of the u, d, s, c, b squarks, 
under the assumption that they are mass degenerate. 
The D0 and CDF Collaborations have published limits on degenerate 
squarks~\cite{d0ds,cdfds}. 
 These limits are outside the LEP2 kinematic range for the case of 
a light gluino; however
limits from LEP2 are competitive with those from the Tevatron if the gluino
is  heavy.

\section{The ALEPH detector}

A detailed description of the ALEPH detector can be found in Ref.~\cite{Alnim},
and an account of its performance as well as a description of the
standard analysis algorithms can be found in Ref.~\cite{Alperf}.
Only a brief overview is given here.

Charged particles are detected in a magnetic spectrometer
consisting of a silicon vertex detector (VDET), a drift chamber (ITC)
and a time projection chamber (TPC), all immersed in a
1.5~T axial magnetic field provided by a superconducting solenoid.
The VDET consists of two cylindrical layers of silicon microstrip detectors;
 it performs very precise measurements of the impact parameter in space 
 thus allowing powerful short-lifetime particle tags, as described  
in Ref.~\cite{Rb1}.
Between the TPC and the coil, a highly granular electromagnetic
calorimeter (ECAL) is used to identify electrons and photons and to
measure their energies. Surrounding the ECAL is the return yoke for the
magnet, which is instrumented with streamer tubes to form the hadron
calorimeter (HCAL). Two layers of external streamer tubes are used 
together with the HCAL to identify muons. 
The region near the beam line is covered by two luminosity calorimeters,
 SICAL and  LCAL, which  provide coverage down to 34~mrad.  
The information obtained from the tracking system is combined
with that from the calorimeters to form a
list of ``energy flow particles''~\cite{Alperf}. These objects serve to
calculate the variables that are used in the analyses described
in Section~3.

\section{The Analyses}

Data collected at 
$\sqrt{s}$ = 181, 182, 183, and 184~GeV have been analysed, 
corresponding to integrated luminosities of 
0.2, 3.9, 51.0, and 1.9~$\rm{pb}^{-1}$, respectively.
Three separate analyses are used to search for the 
processes \mbox{$\mathrm \stop \rightarrow c \neu$},
\mbox{$\mathrm \sbot \rightarrow b \neu$}, and 
$\mathrm \stop \rightarrow b \ell \snu$.
All of these channels are characterised by missing momentum and energy.
The experimental topology depends largely on $\deltm$, the mass difference 
between the $\tilde{\rm{q}}$ and the $\neu$ or $\snu$. When $\deltm$ is large,
there is a substantial amount of energy available for the visible system
and the signal events tend to look like $\ww$, 
$\ewnu$, $\zz$, and $\qqg$ events.
These processes are characterised by high multiplicity and high 
visible mass $M_{\mathrm{vis}}$.
When $\deltm$ is small, the energy available for the visible
system is small and the signal events are therefore similar to $\ggqq$ events.
The process $\ggqq$ is characterised by low multiplicity, 
low $M_{\mathrm{vis}}$, low total transverse momentum $\pt$ 
and the presence of energy near the beam axis. In order to cope with
the different signal topologies and background situations, 
each analysis employs a low $\deltm$ selection and a high $\deltm$
selection.
The values of the analysis cuts are set in an unbiased way following the 
$\bar{N}_{95}$ procedure~\cite{N95}. 
The simulation of the $\ggqq$ background is difficult. As a consequence,
 a safer rejection of this background is ensured by applying tighter
 cuts than would result from the $\bar{N}_{95}$ procedure.

The analyses used to search for evidence of stop and sbottom
production are quite similar to those used at 
$\rts$ = 172~GeV~\cite{ALEPH_stop} with the addition of b tagging in the
 channel $\sbot \to \rm{b} \neu$ 
 to further reject the $\ww$, $\ewnu$, and $\zz$ background. The differences
between the cuts used at $\rts$ = 130--172 GeV and the 
cuts used at $\rts$ = 181--184 GeV are described in detail below.

\subsection{Search for $\stop \to \rm{c} \neu$}

The process $\mathrm e^{+}e^{-} \rightarrow \stop\bar{\stop}$ 
($\mathrm \stop \rightarrow c\neu$)
is characterised by two acoplanar jets and missing mass and energy. 

For the small $\deltm$ selection, only the
thrust and the  visible mass cuts needed adjustments.
 The thrust is required to be
less than 0.915 to reduce further the low-multiplicity $\ggqq$ background.
The lower cut on the visible mass $\mvis$, which is effective against 
$\ggqq$ background, 
  depends upon the mass difference of the 
signal considered. Since signal events with small $\deltm$ tend to have
smaller values of $\mvis$, the optimal value of this cut 
decreases as $\deltm$ decreases. The visible mass is required to be in
 excess of 4~$\gev$. This cut is raised to 7.5~$\gev$ for
 $\deltm$ $>$ 7~$\gev$.

For large $\deltm$, the selection is quite similar to the selection
at $\rts$ = 130--172~GeV. However, a few changes have been made in
order to confront the increased level of background from $\ww$, $\ewnu$,
and $\zz$ that results from the increased luminosity.
The $\thscat$ variable is used to reduce  background from $\ewnu$.
This variable was introduced in Ref.~\cite{ALEPH_stop} as
a means of eliminating $\ggqq$ background. 
Assuming that one of the incoming electrons is scattered while 
the other one continues  undeflected, the polar angle of the
scattered electron, $\thscat$, can then be calculated from the missing
transverse momentum $\pt$.
 This variable can also be interpreted in
the context of $\ewnu$ background. The final state of this background
typically includes an electron which goes down the beampipe and a
neutrino which is within the detector acceptance. In this case, $\thscat$ 
is an estimate of the neutrino polar angle which   
 tends to be large for  the $\ewnu$  background. For
the signal process, $\thscat$ tends to be smaller, as long as $\deltm$ 
is not too large. The optimal value of this cut 
for a hypothesis of $\deltm$ $<$ 35~$\gev$ 
 is  $\thscat$ $<$ $60^{\circ}$, while for 
a hypothesis of $\deltm$ $>$ 35~$\gev$ the cut is not applied.

Background from $\ww$, $\ewnu$, and $\zz$ is also addressed by tightening
the $\mvis$/$\rts$ cut. This cut also depends on the $\deltm$ of the
signal being considered. A hypothesis on $\deltm$=15~$\gev$ gives an
optimal cut of $\mvis$/$\rts$ $<$ 0.25, while a hypothesis on $\deltm$=35~$\gev$ 
 gives an optimal cut of $\mvis$/$\rts$ $<$ 0.33. Finally, the 
optimal cut for all $\deltm$ greater than 50~$\gev$ is $\mvis$/$\rts$ $<$ 0.35.

\subsubsection*{Selection efficiency and background}

According to the $\bar{N}_{95}$ procedure, 
  the low $\deltm$ selection is used
for $\deltm$ $<$ 15~$\gev$,
 while for $\deltm$ $\geq$ 15~$\gev$, the high $\deltm$ selection is used.
The changeover occurs at a larger $\deltm$ value than for 
$\rts$ = 130--172~GeV, where it was  $\deltm$ = 10~$\gev$, due to the
larger contamination in the high $\deltm$ selection.
 
\begin{figure}[tcb]
\begin{center}
\begin{tabular}{cc}
\epsfig{file=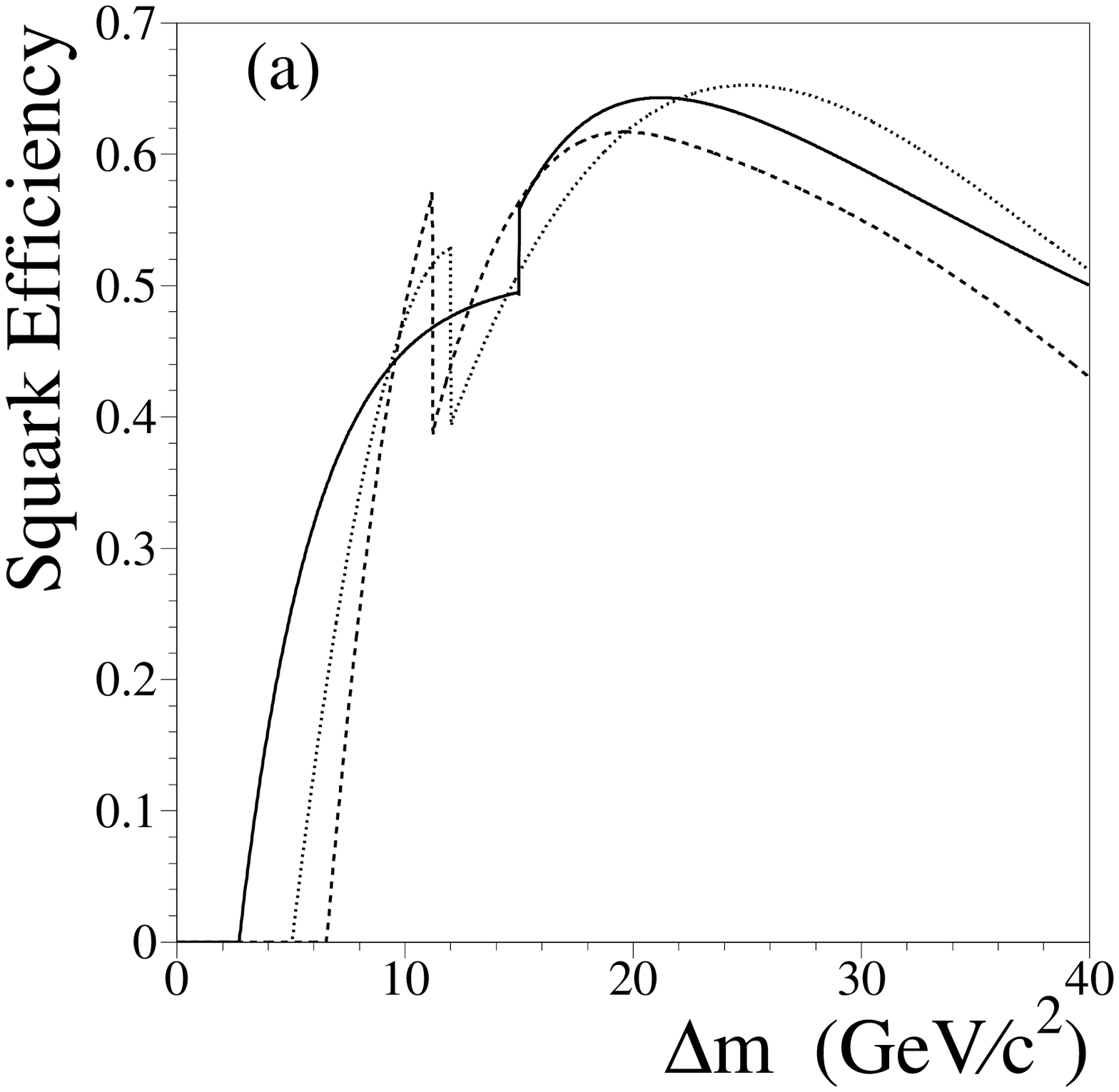,width=0.5\textwidth} &
\epsfig{file=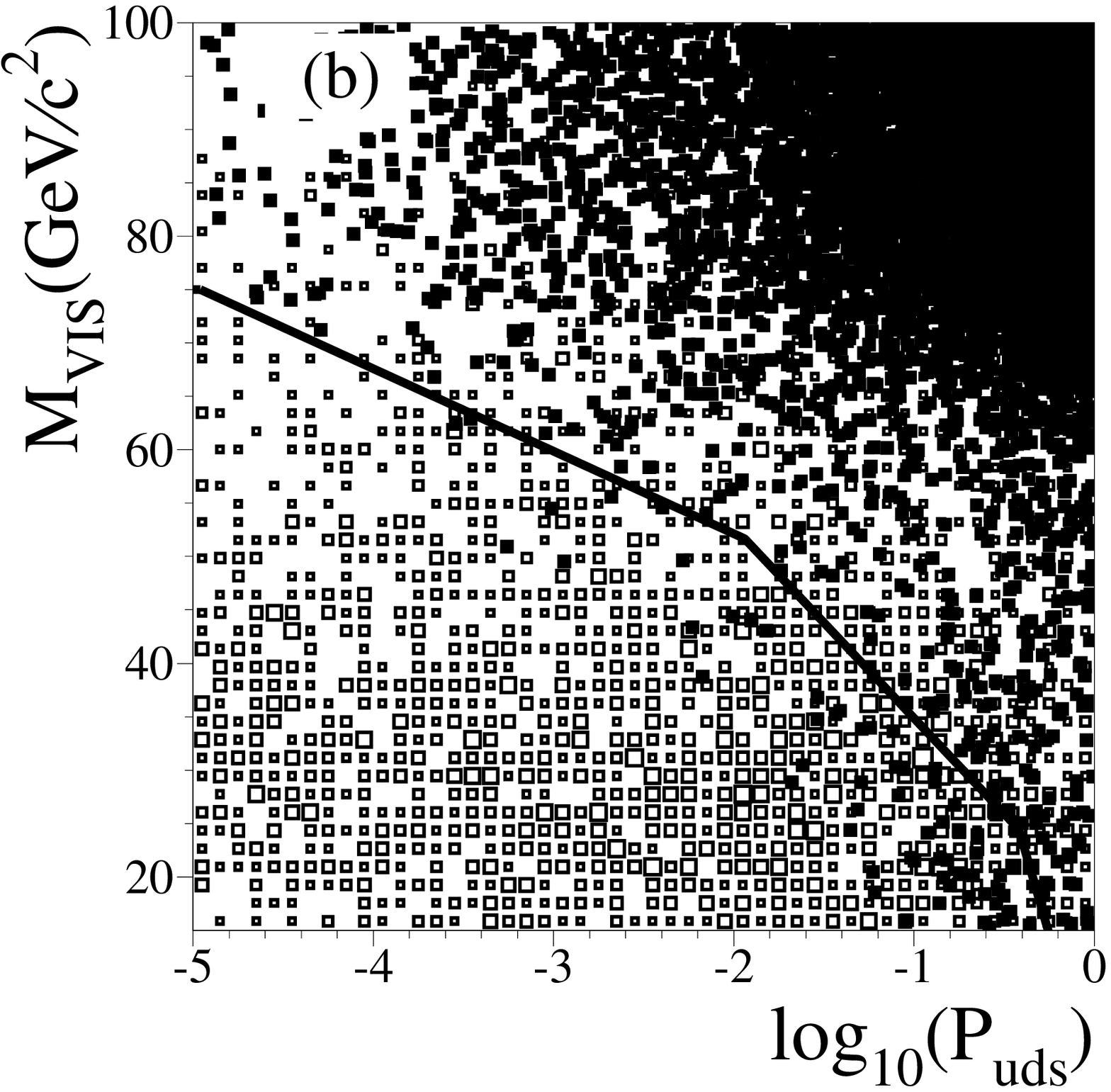,width=0.5\textwidth}
\end{tabular}
\end{center}
\caption{\rm (a) Efficiencies as a function of $\deltm$:
 for an 80 $\gev$ stop decaying as
$\stop \rightarrow \mathrm{c}\neu$ (solid curve), for an
  80~$\gev$ stop decaying as
$\stop \rightarrow \mathrm{b}\ell\tilde{\nu}$  (dashed curve),
 and an 80 $\gev$ sbottom  decaying as
$\sbot \rightarrow \mathrm{b}\neu$ (dotted curve).
 (b) Two-dimensional cut in the plane $\mvis$-log$_{10}$(P$_{\rm{uds}}$).
 Signal (open squares) and background (solid squares) distributions 
 are also shown.
\label{steff}}
\end{figure}

The $\stop \to \rm{c} \neu$ efficiencies are shown in Figure~\ref{steff}a; the discontinuity
at $\deltm$ = 15~$\gev$ is due to the switching between the low and high
$\deltm$ selections.

The background to the low $\deltm$ selection is dominated by
$\ggqq$ and $\ggtt$ and has a total expectation of
1.5~events \mbox{($\sim$ 30 fb)} for the looser value of the
lower $\mvis$ cut. 
For the high $\deltm$ selection, the background is dominated by
$\ww$, $\ewnu$, $\zz$, and $\qqg$. If the 
upper cut on $\thscat$ is not applied and the loosest value of the
upper cut on $\mvis$ is applied, the total 
background expectation for the high $\deltm$  selection
is 3.5 events \mbox{($\sim$ 60 fb)}.

\subsection{ Search for $\sbot \to \rm{b} \chi$ }

The experimental topology of the 
$\mathrm e^{+}e^{-} \to \tilde{b}\bar{\tilde{b}}$ 
($\sbot \to \rm{b}  \chi$)   process is characterised by two acoplanar 
b jets and missing mass and energy.
Both the low and high $\Delta m$ selections use the same selection criteria 
against $\gamma\gamma\rightarrow \rm{q} \bar{\rm{q}} $ background as at lower 
energies~\cite{ALEPH_stop} with cuts rescaled to the centre-of-mass energy 
when appropriate.  
Only the cuts against WW, ${\rm Z\gamma^* }$, and
 We$\nu$ have been reoptimized.
 Most of the cuts are similar to  
those used for the $\stop \to \rm{c} \chi$ process;
 here only the differences  will be described.    

For the low $\Delta m$ selection,   
the visible mass of the event is required to be greater than 7.5  
GeV/c$^{2}$. In this channel the b quark in the final state 
produces a visible mass higher than in the 
$\stop \to \rm{c} \chi$ channel.  
 
For the high $\Delta m$ selection,
the level of    WW, $\rm{ Z\gamma^* }$ and  We$\nu$ background 
 was  reduced by  taking advantage of the lifetime  content of 
the $\sbot \rightarrow \rm{b} \chi$  topology. 
The b quark events were tagged by
using the b-tag event probability $({\rm P_{uds}})$ described in ~\cite{Rb1}.
 Since this probability  depends largely on the 
b jet boost, events are  more b-like as the event visible mass  increases.
A two-dimensional cut is applied in the
 $\mvis$ vs $\log_{10}({\rm P_{uds}})$ plane (Figure~\ref{steff}b); 
 starting from a loose value of the b-tag
 cut when the visible mass is  low, the cut becomes  tighter 
 for larger values of visible mass. 

\subsubsection*{Selection efficiency and background }

According to the $\bar{N}_{95}$  procedure,  
for $\Delta m <$ 12 GeV/$c^{2} $ the low
$\Delta m $ selection is used while for  $\Delta m >$ 12 GeV/$c^{2} $
the high $\Delta m $ selection is used. 
The efficiency  is  shown in  Figure~\ref{steff}a,  
as a function of $\Delta m$ for $M_{\sbot}$=80 GeV/$c^{2}$.
The total  background expectation for the low  
$\Delta m $ selection is 1.1 events
(20 fb), and dominated by $ \gamma\gamma\rightarrow \rm{q\bar{q}}$. 
For the high  $\Delta m $ selection the  WW, ${\rm Z\gamma^*} $, and We$\nu$  
background is highly suppressed by b tagging and the total 
background expectation  is  0.6 events (10 fb).

\subsection{Search for $\stop \to \rm{b} \ell \snu$}

The experimental signature for
$\mathrm e^{+}e^{-} \rightarrow \stop\bar{\stop}$ 
($\stop \to \rm{b} \ell \snu$) is
two acoplanar jets plus two leptons with missing momentum.

For the low $\deltm$ selection, the $\pt$ cut used at 172 GeV
      is reinforced by requiring that the $\pt$ calculated without
      neutral hadrons and the $\pt$ calculated with only charged tracks
      both be greater than 0.75\%$\sqrt{s}$.


For the large $\deltm$ selection, the cuts are optimised in
order to confront the increased rate of $\ww$ and $\zz$ backgrounds at  
183~GeV. Only two cut values are changed with respect to their values at
172~GeV. First, the upper cut on the hadronic mass is tightened: the 
hadronic mass is required to be less than 25\%$\rts$ if one 
lepton is identified and less than 20\%$\rts$ if more than one lepton is  
identified. Additionally, the 
cut on the leading lepton isolation is reinforced: 
the energy in a $30^{\circ}$ cone around the direction of the electron 
or muon momentum must be  smaller than 
2.7 times the electron or muon energy.

\subsubsection*{Selection efficiency and background}
The  combination of the two selections is chosen  
according to the $\bar{N}_{95}$ procedure. 
For $\deltm$ $<$ 11~$\gev$, the logical OR of the two selections
is used: the high $\deltm$ selection helps to recover efficiency 
 while leaving the background level unchanged.
For $\deltm$ $\geq$ 11~$\gev$, only the high $\deltm$
selection is used. This is in contrast to the situation
for the 130--172 GeV data; because the background was  low
for both the low and the high $\deltm$ selections, the OR
of the two selections was optimal for all $\deltm$.

The  $\stop \to \rm{b} \ell \snu$ selection efficiencies
are given in Figure~\ref{steff}a. The effect of switching from
the OR of the two selections to the high $\deltm$ selection
by itself can be seen at $\deltm$ = 11~$\gev$.  
The background to the low $\deltm$ selection is dominated by
$\ggqq$ and has a total expectation of
 0.8~events \mbox{($\sim$ 14 fb)}. 
For the high $\deltm$ selection, the very low expected background    
 (0.1 events expected or 
 ${\sim} 2 \,{\rm fb}$) is dominated by $\ww$ events.

\subsection  {Systematic uncertainties}
 
The systematic uncertainties on the $\stop$ and $\sbot$ selection 
efficiencies originating from the physical processes in the Monte Carlo 
simulation as well as those related to detector effects are evaluated
 following the procedure described in Reference~\cite{ALEPH_stop}.
The relative uncertainty on the selection efficiency in the case of
 $\stop \to \rm{c} \chi$  is 
  13\% for low $\Delta m$ and  6\% for  high $\Delta m$;
 in the case of
  $\stop \to \rm{b} \ell \snu$ it is
 16\% for  low $\Delta m$ and  6\% for  high $\Delta m$; 
 for  the $\sbot \to \rm{b} \chi$ channel it is
 12\% for  low $\Delta m$ and  6\% for  high $\Delta m$.
These errors are dominated by the uncertainties on the simulation 
 of $\stop$ and $\sbot$ production and decay.
An additional source of systematic error for the high $\Delta m$
  $\sbot \to \rm{b} \chi$ selection efficiency 
derives from the uncertainty on the b-tagging.
This systematic uncertainty has been studied by measuring 
$ R_{\rm{b}}$
 as a function of the b-tag cut   
 in the calibration data collected at the Z peak during the 1997 run.
  The total uncertainty on the efficiency for the high $\Delta m$
 $\sbot \to \rm{b} \chi$ selection is 7\%.
The systematic uncertainties are included  in the final result following the 
method described in \cite{syse}.

\section{Results}

A total of five events are selected by the $\stop \to \rm{c} \neu$ 
analysis,  four by the high $\deltm$ selection and one by the
low $\deltm$ selection. This is consistent with the 3.5 events
expected in the high $\deltm$ selection and the 1.5 events expected
in the low $\deltm$ selection. The kinematic properties of the high
$\deltm$ events are all consistent with  $\zz$, $\ww$,
or $\ewnu$, while the kinematic properties of the low $\deltm$ event 
suggest the process $\ggqq$.

A single event is selected by the $\sbot \to \rm{b} \neu$ analysis. 
This event, which is found by the high $\deltm$ selection, 
is also found by the $\stop \to \rm{c} \neu$ high $\deltm$ selection.
The three high $\deltm$ $\stop$ candidates not selected by the 
sbottom analysis are all rejected by the b-tag. The number of events
selected in the data is consistent with the  expectation from
background processes
 (1.1  from the low and 0.6 from the high $\deltm$ selections).

A single event is also selected by the $\stop \to \rm{b} \ell \snu$
analysis. The event is found by the low $\deltm$ selection, and
is consistent with $\ggqq$ production. The total
of one event selected is consistent with the 0.9 events that are
expected from background processes 
(0.8  from the low and 0.1 from the high $\deltm$ selections).

\begin{figure}[p]
\begin{center}
\begin{tabular}{cc}
\epsfig{file=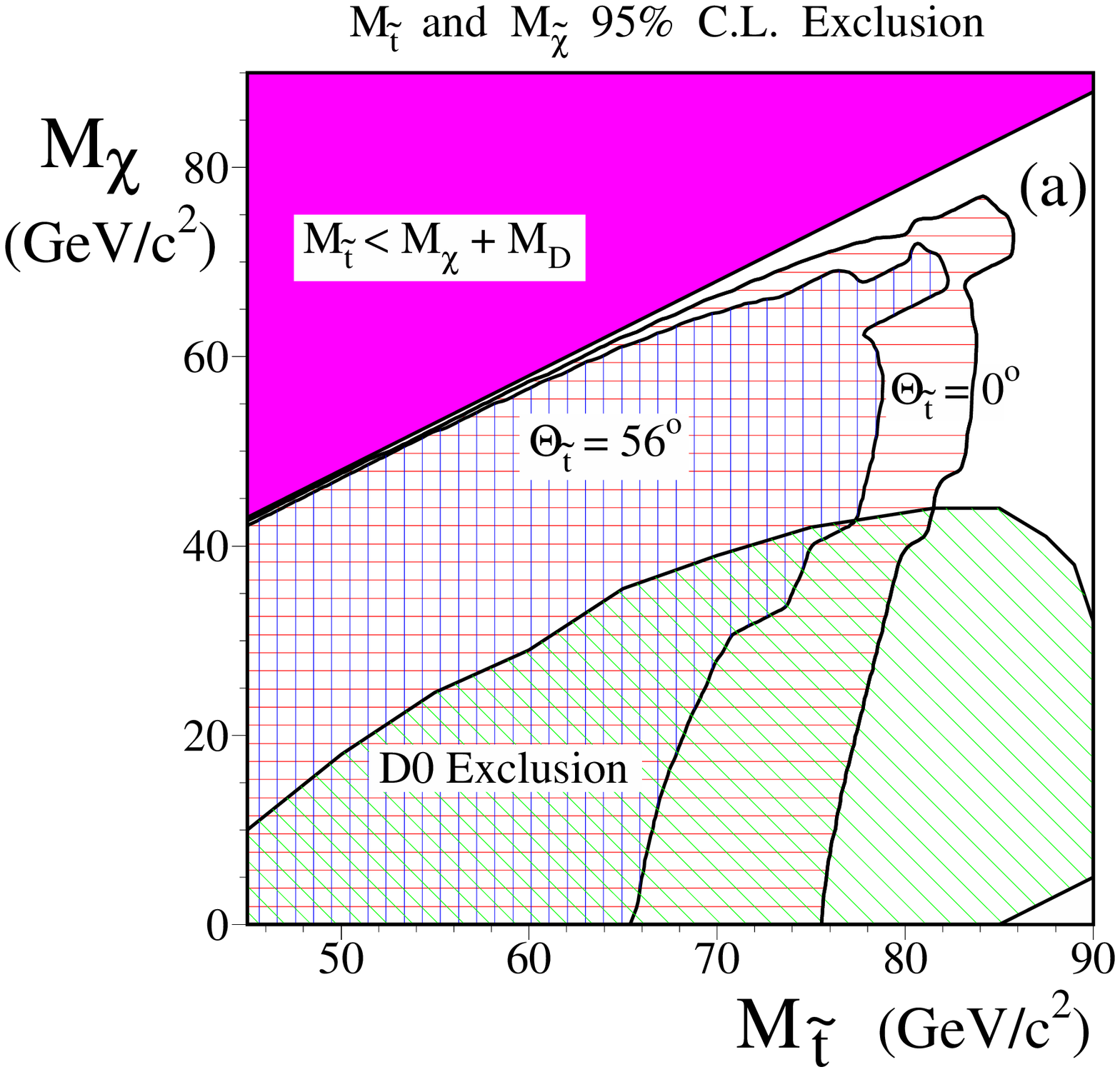,width=0.46\textwidth} &
\epsfig{file=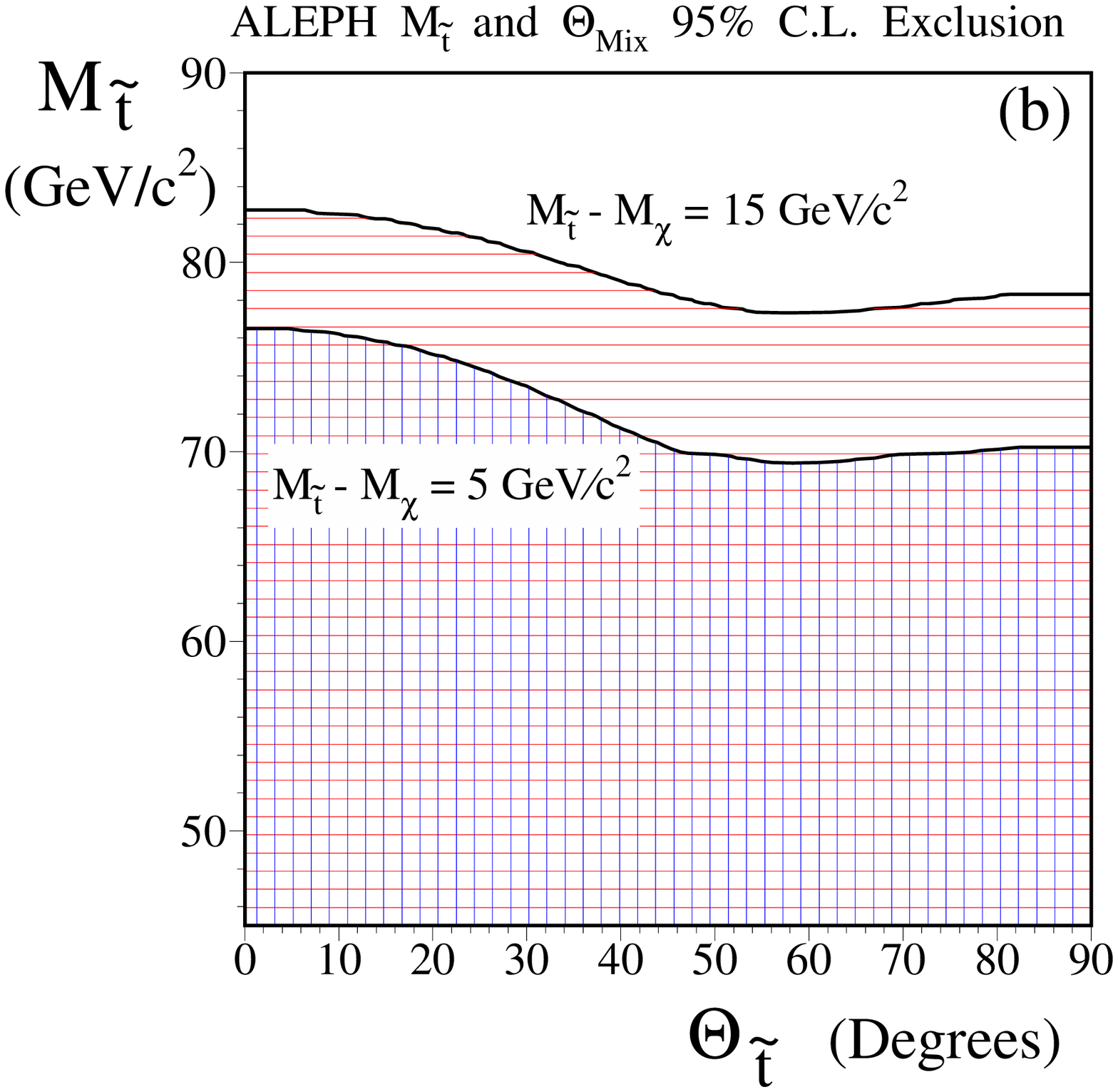,width=0.46\textwidth}
\end{tabular}
\end{center}
\caption{\rm Excluded regions for $\stop \rightarrow \mathrm{c}\neu$. 
(a) Excluded region in the $m_{\neu}$ vs 
$m_{\stop}$ plane; the region excluded by the
D0 collaboration is also indicated. (b) Excluded region in the  
$m_{\stop}$ vs $\mathrm \thetamix$ plane. 
 In (a), the excluded regions are given for $\mathrm \thetamix$=$0^{\circ}$,
 corresponding to  maximum $\stop$-Z coupling, and for
$\mathrm \thetamix$=$56^{\circ}$, corresponding to  minimum $\stop$-Z coupling.
\label{stchi}}
\end{figure}

\begin{figure}[p]
\begin{center}
\begin{tabular}{cc}
\epsfig{file=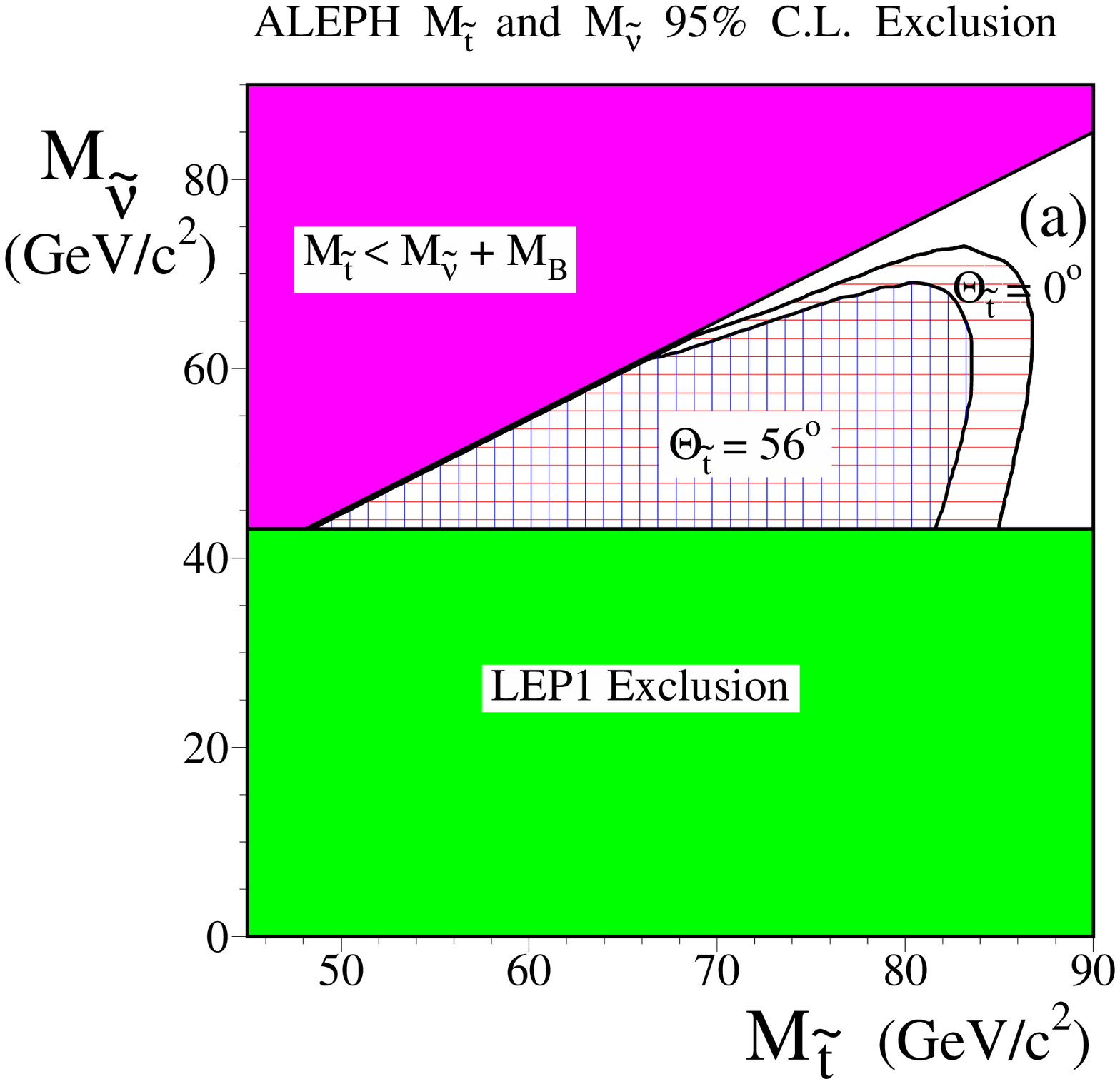,width=0.46\textwidth} &
\epsfig{file=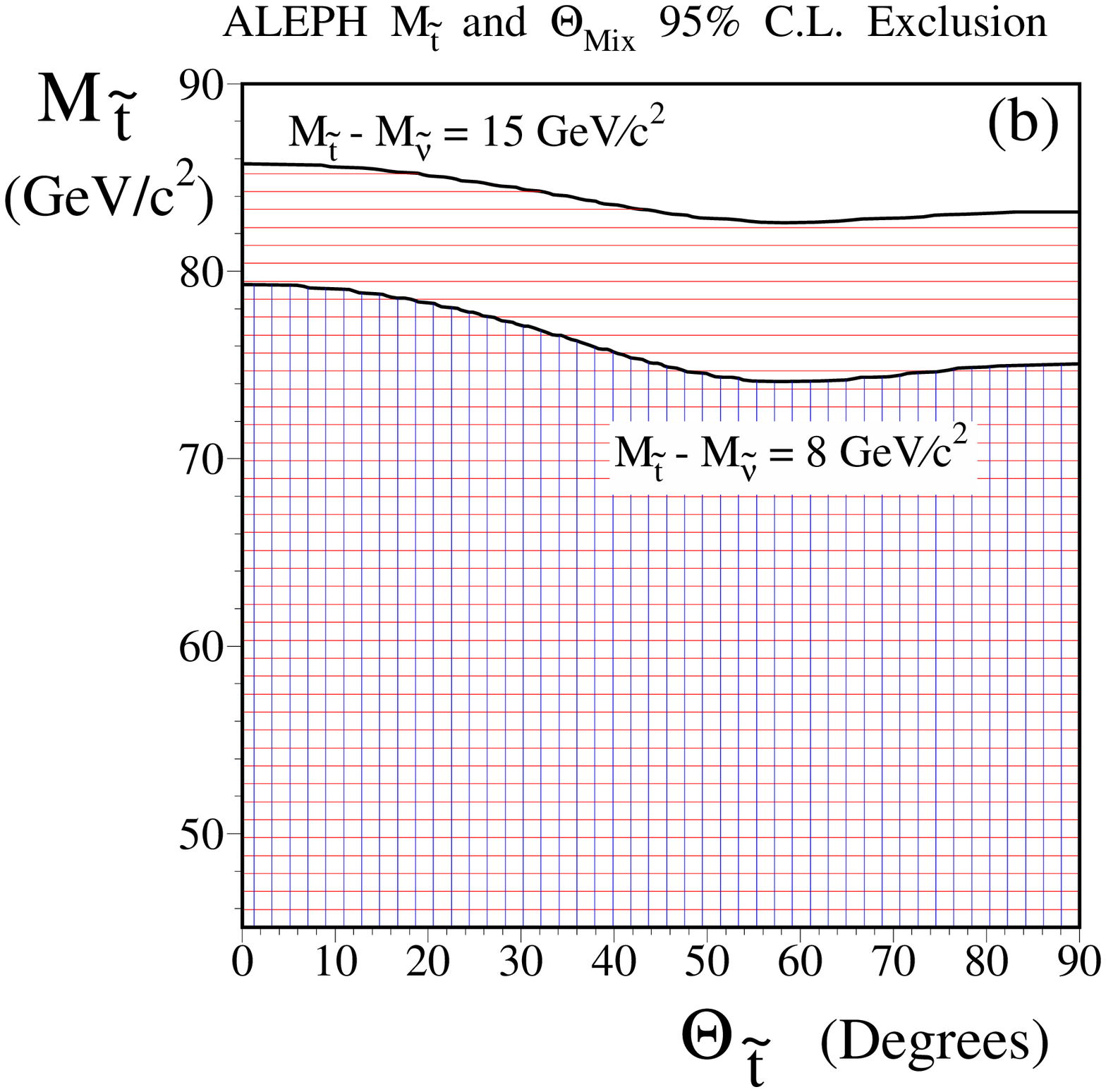,width=0.46\textwidth}
\end{tabular}
\end{center}
\caption{\rm Excluded regions for  
$\stop \rightarrow \mathrm{b}\ell\tilde{\nu}$ 
(equal branching fractions for the $\stop$ decay to $\mathrm{e}$,
$\mu$, and $\tau$ are assumed).
(a) Excluded region in the $m_{\tilde{\nu}}$ vs 
$m_{\stop}$ plane. (b) Excluded region in 
the $m_{\stop}$ vs 
$\mathrm \thetamix$ plane.   
In (a), the excluded regions are given for $\mathrm \thetamix$=$0^{\circ}$,
 corresponding to  maximum $\stop$-Z coupling, and for 
$\mathrm \thetamix$=$56^{\circ}$, corresponding to  minimum
 $\stop$-Z coupling. Also 
shown in (a) is the region excluded from LEP1 data, i.e., the $\snu$ mass
 limit obtained from the measurement of the Z lineshape.
\label{stblv}}
\end{figure}

Since no evidence for the production of $\stop$ or $\sbot$
 is found, it is appropriate to set lower limits on their
masses. The limits are extracted without background subtraction.    
Figures~\ref{stchi}a and \ref{stchi}b give the 95\% C.L.
excluded regions for the channel $\mathrm \stop \rightarrow c\neu$.
For this channel, the $\thetamix$-independent
lower limit on $m_{\stop}$ is 74~$\gev$, assuming a mass difference between 
the $\stop$ and the $\neu$ of 10--40~$\gev$, corresponding to a large
 part of the region not excluded by the D0 search.
Figures~\ref{stblv}a and \ref{stblv}b  give the excluded
regions for the $\mathrm \stop \rightarrow b\ell\widetilde{\nu}$ channel,
assuming equal branching ratios for the $\stop$ decay 
to $\mathrm{e}$, $\mu$ and $\tau$. In this case, 
the $\thetamix$-independent lower limit on $m_{\stop}$ is 82~$\gev$, 
assuming a mass difference between the $\stop$ and the $\snu$ of at 
least 10~$\gev$, and using also the LEP1 exclusion on the sneutrino mass.

\begin{figure}[p]
\begin{center}
\begin{tabular}{cc}
\epsfig{file=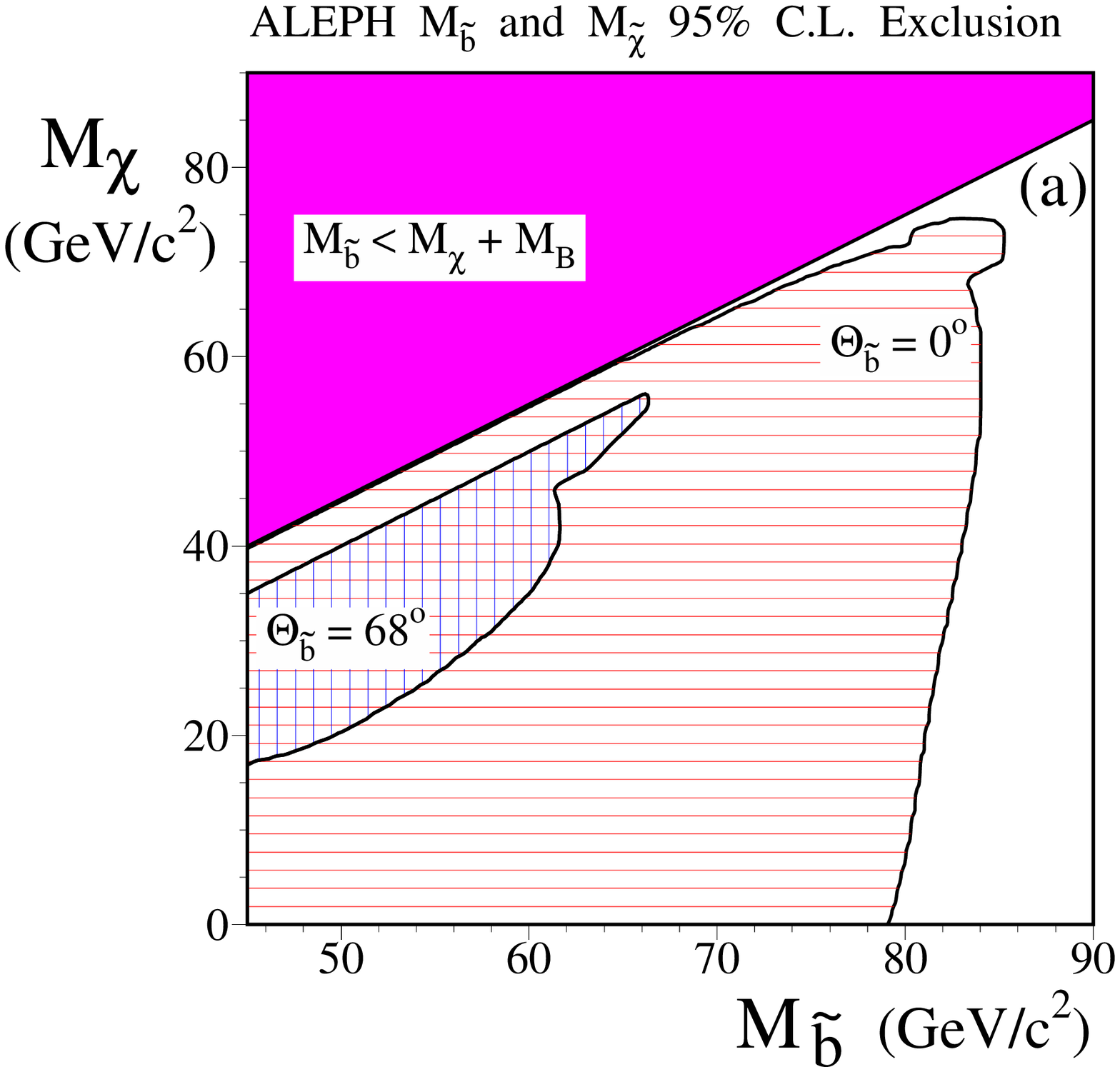,width=0.46\textwidth} &
\epsfig{file=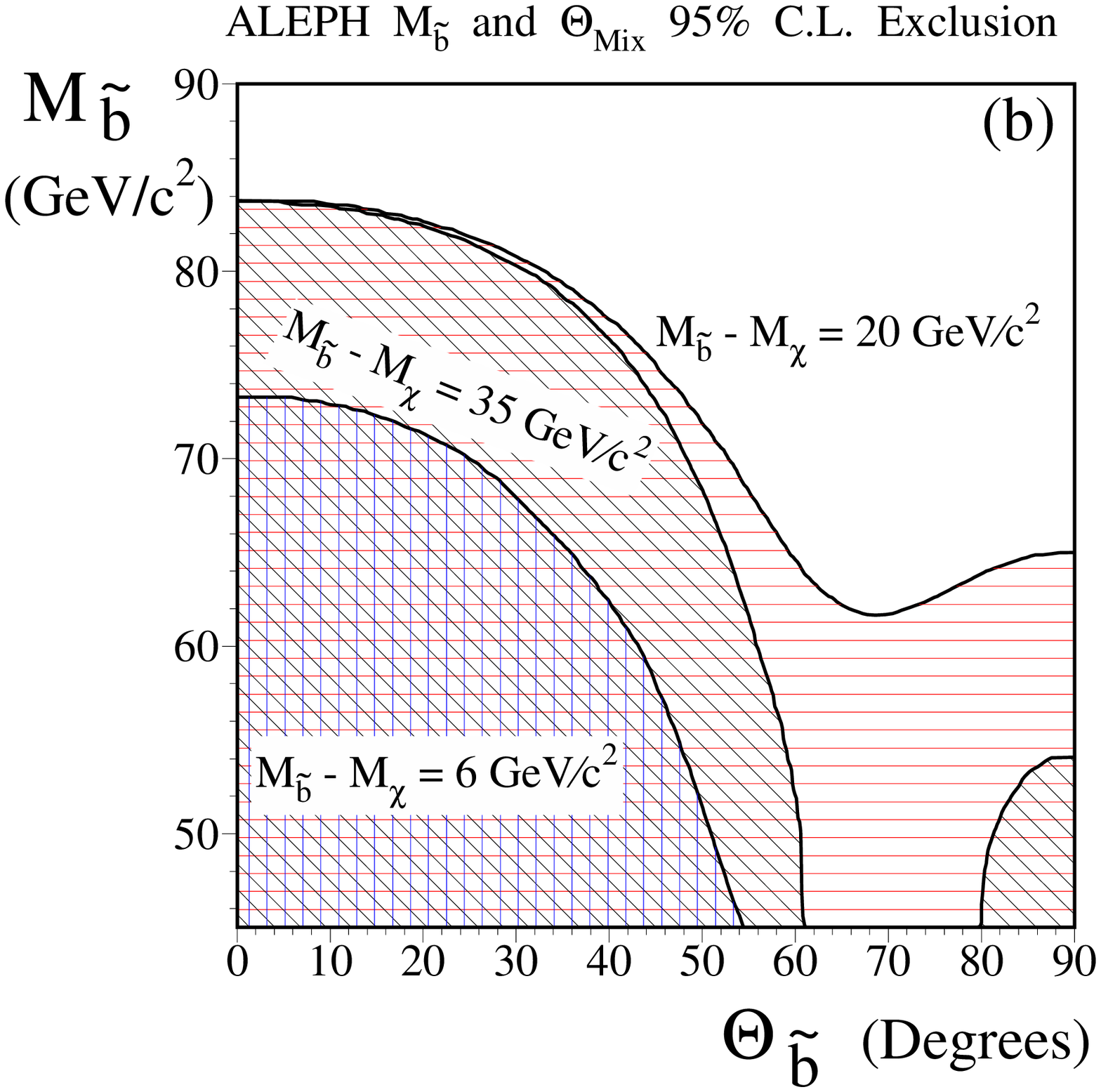,width=0.46\textwidth}
\end{tabular}
\end{center}
\caption{\rm Excluded regions for $\sbot \rightarrow
\mathrm{b}\neu$. (a) Excluded region in the $m_{\neu}$ vs 
$m_{\sbot}$ plane. 
(b) Excluded region in the $m_{\sbot}$ vs 
$\mathrm \thetab$ plane. 
In (a), the excluded regions are given for $\mathrm \thetab$=$0^{\circ}$,
 corresponding to  maximum $\sbot$-Z coupling, and for 
$\mathrm \thetab$=$68^{\circ}$, corresponding to  minimum $\sbot$-Z coupling. 
\label{sbotchi}}
\end{figure}

Figures~\ref{sbotchi}a and \ref{sbotchi}b 
give the excluded regions for the
$\sbot$ in the decay channel $\mathrm \sbot \rightarrow b\neu$. 
A lower limit of 79~$\gev$ is set on $m_{\sbot}$, assuming that
$\thetab$ is $0^{\circ}$ and that the mass difference between the 
$\sbot$ and the $\neu$ is at least 10~$\gev$.
Figure~\ref{sbotchi}b shows that only a restricted 
region is excluded when $\thetab$ =$68^{\circ}$.
When decoupling from the Z occurs, sbottoms can only be produced
through photon exchange and the cross section for the $\sbot$ 
(charge $-1/3$) is four times lower than the cross section for the $\stop$
(charge $+2/3$).

\subsection*{Limits on degenerate squarks}

Here the decay $\tilde{\rm{q}} \to \rm{q}\neu$ is assumed to be dominant.
It has a topology similar to that of $\stop \to \rm{c}\neu$. 
 The $\stop \to \rm{c}\neu$ analysis can therefore
be used to search for generic squark production. In order to check the
efficiency of the $\stop \to \rm{c}\neu$ selection when it is applied
to degenerate squark production, samples of the process
$\tilde{\rm{q}} \to \rm{q}\neu$ were generated and run through the
full ALEPH detector simulation. As expected, the selection efficiency 
for these samples is similar to the selection efficiency for the 
corresponding $\stop \to \rm{c}\neu$ samples. The 
$\tilde{\rm{q}} \to \rm{q}\neu$ efficiencies were then parametrised as
a function of squark mass and $\deltm$, and this parametrization is 
used to set the limits on generic squark production.
For q = u, d, s, or c, the mixing between
$\rm{\tilde{q}_{R}}$ and $\rm{\tilde{q}_{L}}$ is expected to be negligible.
 The 
mixing between $\rm{\tilde{b}_{R}}$ and $\rm{\tilde{b}_{L}}$ is also assumed
to be negligible in the case that the sbottoms are mass
degenerate with the partners of the four lightest quarks. 

\begin{figure}[p]
\begin{center}
\begin{tabular}{cc}
\epsfig{file=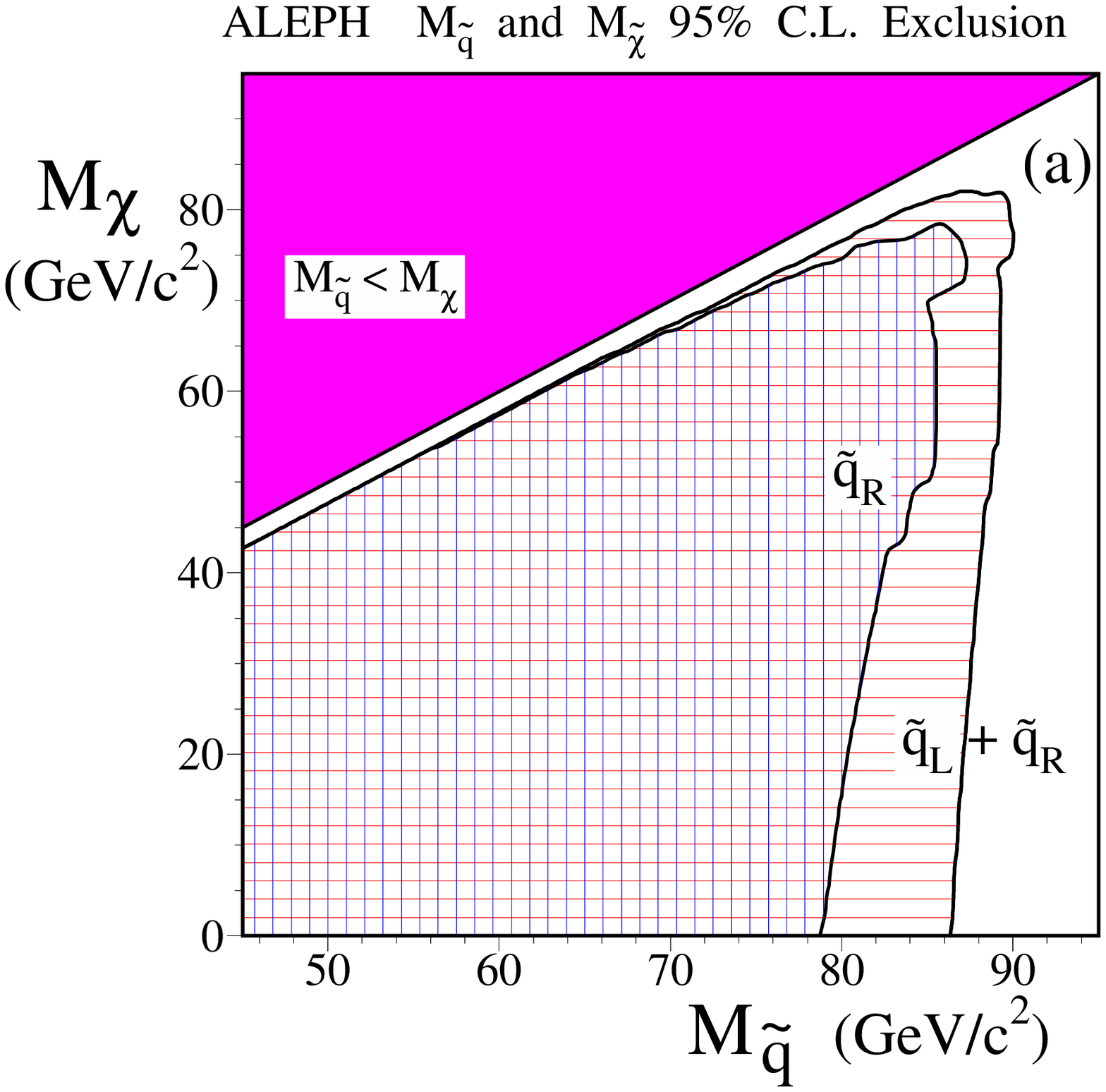,width=0.46\textwidth} &
\epsfig{file=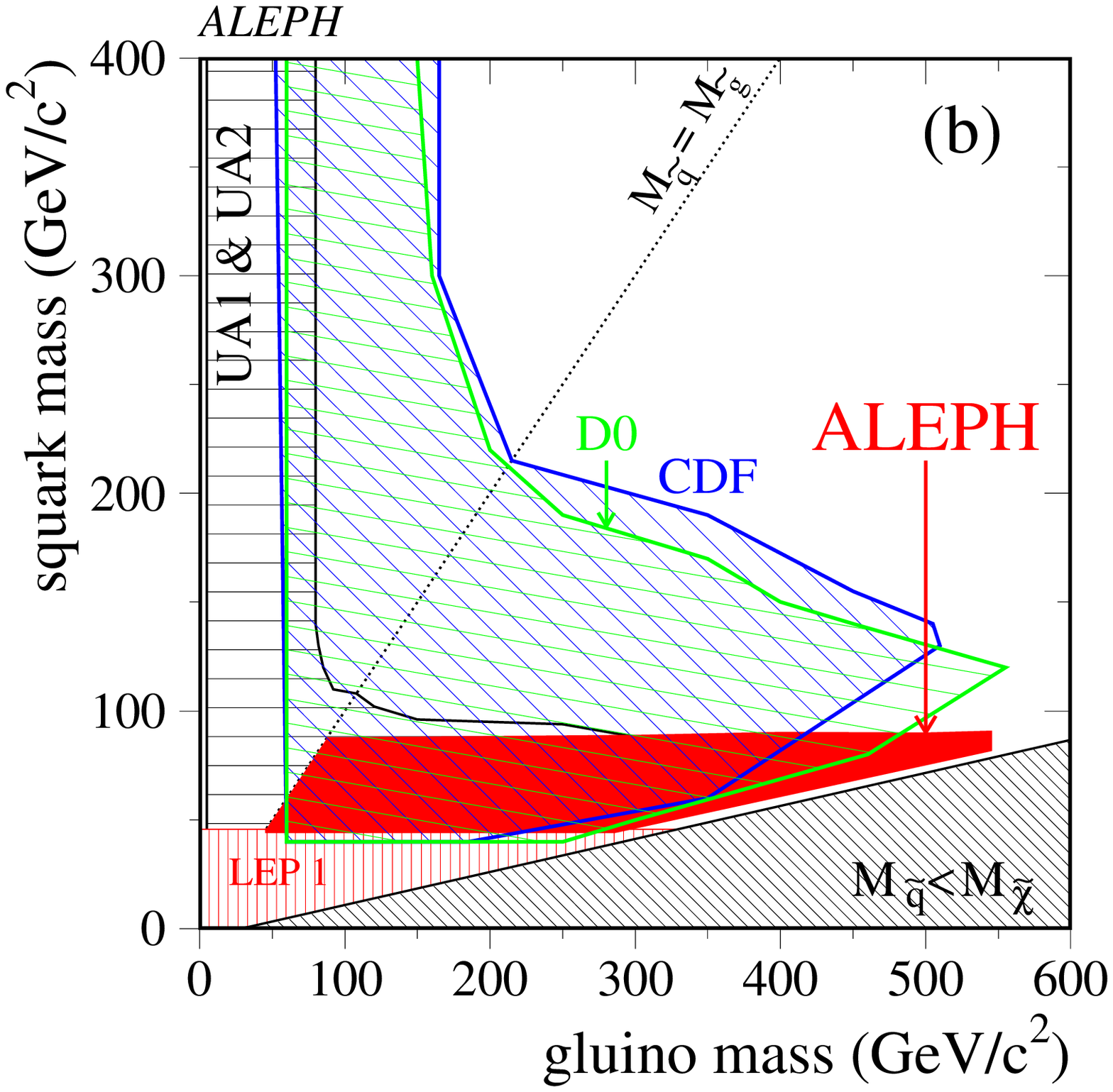,width=0.46\textwidth}
\end{tabular}
\end{center}
\caption{\rm 
(a) Excluded region for the
supersymmetric partners of the u, d, s, c, and b quarks, 
 assuming the squarks to be
degenerate in mass. In (a), two curves are given: one curve
assumes that only ${\rm\tilde{q}}_{\rm{ R}}$ is accessible at LEP2 energies, while
the other curve assumes that ${\rm\tilde{q}}_{\rm{ R}}$ and
 ${\rm\tilde{q}}_{\rm{L}}$
 are both
accessible at LEP2 energies. (b) The ALEPH result for five degenerate flavours,
${\rm \tilde{q}}_{\rm{R}}$ and ${\rm\tilde{q}}_{\rm{L}}$ production,
 shown in the gluino-squark
mass plane for tan$\beta$ = 4 and $\mu$ = $-400 \gev$.
 This result excludes a small region not excluded by CDF and D0.
\label{degsq}}
\end{figure}

Figure~\ref{degsq}a shows the exclusion curves assuming five degenerate
squark flavours.
 In the more conservative curve, only
$\rm{\tilde{q}_{R}\bar{\tilde{q}}_{R}}$ production is allowed, while in the
other curve, both $\rm{\tilde{q}_{R}\bar{\tilde{q}}_{R}}$ and
$\rm{\tilde{q}_{L}\bar{\tilde{q}}_{L}}$ production are allowed, assuming that 
$\rm{\tilde{q}_{L}}$ and $\rm{\tilde{q}_{R}}$  are mass degenerate. For 
these curves the efficiency parametrisation developed for the dedicated
sbottom search has been applied to the processes 
$\mathrm e^{+}e^{-} \to \tilde{b}_{R}\bar{\tilde{b}}_{R}$ and
$\mathrm e^{+}e^{-} \to \tilde{b}_{L}\bar{\tilde{b}}_{L}$.

Searches for degenerate squarks have been performed by CDF and D0 at the 
Tevatron; the resulting limits in the gluino-squark mass plane
 are shown in Figure~\ref{degsq}b. While these experiments can exclude quite 
an extensive region of this plane, there is an uncovered region
 in the exclusion for large gluino masses.

In the MSSM, when GUT relation are assumed, the neutralino mass can be 
 related to the gluino mass once the values of $\mu$ and tan$\beta$
 are fixed. Therefore, ALEPH limits in the squark-neutralino 
 mass plane can be translated to the gluino-squark mass plane.
 The ALEPH results are shown in Figure~\ref{degsq}b assuming 
 the values of $\mu$ and tan$\beta$ used by CDF 
(tan$\beta$ = 4 and $\mu$ = $-400~\gev$).
This exclusion is affected only slightly if the D0 set of values 
(tan$\beta$ = 2 and $\mu$ = $-250~\gev$) is used instead.

 The limit on $m_{\rm{\tilde{q}}}$ is at least
87~$\gev$ up to \mbox{$m_{\tilde{{\rm g}}}$ $\sim$  545~$\gev$} (535 $\gev$ if
 D0 values are used). At this point,
$m_{\neu}$ = 82~$\gev$ and the mass difference between 
$\rm{\tilde{q}}$ and $\neu$ is only 5~$\gev$. 
Beyond these points, the $\stop \to \rm{c} \neu$
analysis is no longer sensitive to the production of degenerate squarks.

\section{Conclusions}
Searches have been performed for scalar quarks  at 
$\rts$ = 183~GeV. Five candidate events are  
observed in the $\mathrm \stop \rightarrow c \neu$ channel,
one in the 
$\mathrm \stop \rightarrow b \ell \tilde{\nu}$ channel, and one in the
$\mathrm \sbot \rightarrow b \neu$ channel. These totals are consistent with 
the expectation from background processes.

\par
A 95\% C.L. limit of $m_{\stop}$ $>$ 74~$\gev$ is obtained from the 
$\mathrm \stop \rightarrow c\neu$  search, independent of the
mixing angle and  for $10 < \Delta m <40~$GeV/$c^2$.  
From the $\mathrm \stop \rightarrow b\ell\widetilde{\nu}$ channel, the 
$\thetamix$-independent limit \mbox{$m_{\stop}$ $>$ 82~$\gev$}
is established, if the mass
difference between the $\stop$ and the $\snu$ is greater than
10~$\gev$ and for equal  branching ratios  of the
$\stop$ into $\mathrm{e}$, $\mu$, and $\tau$.

\par
A limit is also obtained for the $\sbot$ decaying as 
$\mathrm \sbot \rightarrow b\neu$. The limit is $m_{\sbot}$ $>$ 
79~$\gev$ for the supersymmetric partner of the left-handed state of
the bottom quark if the mass difference between
the $\sbot$ and the $\neu$ is greater than 10~$\gev$. 

Finally, limits are also derived for the supersymmetric partners
of the light quarks. Assuming five degenerate flavours and the production
of both ``left-handed'' and ``right-handed'' squarks, a limit of 
$m_{\rm{\tilde{q}}}$ 
$>$ 87~$\gev$ is set. This limit is valid for $\deltm$ $>$ 5~$\gev$. Using the
GUT relations, for tan$\beta$ = 4 and $\mu$ = $-400$ $\gev$, 
the limit of $m_{\rm{\tilde{q}}}$ $>$ 87~$\gev$
 is valid for gluino masses smaller than 545~$\gev$.

\section{Acknowledgements}
We wish to congratulate our colleagues from the accelerator
divisions for the continued successful operation of LEP at high energies. 
We would also like to express our gratitude to the engineers
and support people at our home institutes without whom this work
would not have been possible. Those of us from non-member states
wish to thank CERN for its hospitality and support.

\newpage

\newpage

\end{document}